\documentclass[aps,prb,epsf,twocolumn,showpacs,preprintnumbers]{revtex4}
\usepackage{graphics}
\usepackage{graphicx}% Include figure files
\usepackage{dcolumn} % Align table columns on decimal point
\usepackage{bm}
\usepackage{epsfig}
\usepackage{float}
\begin{document}

\title{Electronic and magnetic properties of K$_{2}$CuP$_{2}$O$_{7}$ -
a model S=1/2 Heisenberg chain system}

\author{R. Nath}
\affiliation{Max Planck Institut f\"{u}r Chemische Physik fester Stoffe,
N\"{o}thnitzer Str. 40, 01187 Dresden, Germany}

\author{Deepa Kasinathan}
\affiliation{Max Planck Institut f\"{u}r Chemische Physik fester Stoffe,
N\"{o}thnitzer Str. 40, 01187 Dresden, Germany}

\author{H. Rosner}
\email{rosner@cpfs.mpg.de}
\affiliation{Max Planck Institut f\"{u}r Chemische Physik fester Stoffe,
N\"{o}thnitzer Str. 40, 01187 Dresden, Germany}

\author{M. Baenitz}
\affiliation{Max Planck Institut f\"{u}r Chemische Physik fester Stoffe,
N\"{o}thnitzer Str. 40, 01187 Dresden, Germany}

\author{C. Geibel}
\affiliation{Max Planck Institut f\"{u}r Chemische Physik fester Stoffe,
N\"{o}thnitzer Str. 40, 01187 Dresden, Germany}

\date{\today }

\begin{abstract}
The electronic and magnetic properties of K$_{2}$CuP$_{2}$O$_{7}$
were investigated by means of susceptibility, specific heat and
$^{31}$P nuclear magnetic resonance (NMR) measurements and by LDA
band structure calculations. The temperature dependence of the NMR
shift $K(T)$ is well described by the $S=\frac{1}{2}$ Heisenberg
antiferromagnetic chain (HAF) model with nearest neighbor exchange
$J_{1}$ $\simeq $ $(141\pm 5)$ K.  The corresponding mapping of an
LDA-derived tight-binding model leads to $J_{1}^{LDA}$ $\simeq$
$196$ K. The spin lattice relaxation rate $1/T_{1}$
%is nearly
%$T$-independent below 30 K, but increases linearly with $T$ above 30
%K,
decreases with temperature below 300 K but becomes nearly
temperature independent between 30 K and 2 K as theoretically
expected for an $S = \frac{1}{2}$ HAF chain. None of the
investigated properties give any evidence for long range magnetic
order above 2K, in agreement with the results of the band structure
calculation, which yield extremely weak exchange to the next nearest
neighbor (NNN) and a very small and frustrated inter-chain exchange.
Thus, K$_{2}$CuP$_{2}$O$_{7}$ seems to be a better realization of a
nearest neighbor $S = \frac{1}{2}$ HAF chain than the compounds
reported so far.
\end{abstract}


\maketitle

\section{\textbf{Introduction}}

One-dimensional (1D) spin systems have attracted considerable
attention because of their intriguing ground states where quantum
fluctuations play a crucial role. Much excitement in their magnetism
has been caused by the theoretical prediction that the integer-spin
chains have an energy gap in the excitation
spectrum,\cite{haldane1983} while the half-integer spin chains have a
gapless excitation spectrum.\cite{bethe31, lieb61} In particular, 1D
$S=\frac{1}{2}$ Heisenberg antiferromagnetic (HAF) systems are
interesting since enhanced quantum fluctuations due to reduced
dimensionality and low spin value impede long range magnetic order
(LRO). In the last decade, theoretical studies on these systems have
achieved a remarkable progress but real material realizations for such
compounds are limited to date.

The $S=\frac{1}{2}$ chains formed via direct linkage of CuO$_{4}$
units can be grouped into two categories: in one group, the chains are
formed by edge-sharing CuO$_{4}$ units while in the other compound
family they are built from corner-sharing CuO$_{4}$ units.
CuGeO$_{3}$ and Li$_{2}$CuO$_{2}$ belong to the former
category\cite{mattheiss1994,neudert1999}, where the nearest neighbor
(NN) interaction $J_{1}$ and the next-nearest neighbor (NNN)
interaction $J_{2}$ have comparable strength ($\frac{|J_{2}|}{|J_{1}|}
\approx$ 1) thereby leading to a strong frustration within the
chain. In such a scenario, various ground states are possible.  On the
contrary, in Sr$_{2}$CuO$_{3}$ which has corner shared CuO$_{4}$ units
$\frac{|J_{2}|}{|J_{1}|}$ $\approx$ $\frac{1}{20}$, which greatly
reduces the in-chain
frustration\cite{ami1995,motoyama1996,rosner1997}.  With $J_{1}$ $\gg$
T$_{N}$, Sr$_{2}$CuO$_{3}$ is a better quasi-1D $S=\frac{1}{2}$ chain
system than CuGeO$_{3}$ and Li$_{2}$CuO$_{2}$.  Recently another
system, Sr$_{2}$Cu(PO$_{4}$)$_{2}$ containing isolated CuO$_{4}$ units
(neither edge- nor corner-shared) was reported to have
$\frac{J_{2}}{J_{1}}$ $\approx$ $\frac{1}{700}$ and a much reduced
$T_{N}$ of 85 mK\cite{nath2005,belik2005,johannes2006,salunke2007}.
The presence of inter-chain coupling (ic) normally provides LRO, while
frustrating ic reduces the tendency to LRO.  Along with $J_{1}$ $\gg$
$T_{N}$, the presence of frustration between the chains strongly
influences the 1D nature of this system, making it a nearly perfect
realization for the 1D NN only HAF chain model.  Unfortunately the
sample quality was not very good, thereby reducing the chains to a
finite length.  The analytical solution by Bethe ansatz\cite{bethe31}
provides a clear picture of the magnetic (thermodynamic) properties of
a very good 1D material. Thus, deviations from these predictions can
be assigned to other degrees of freedom (anisotropy,
Dzyaloshinskii-Moriya (DM) and other spin-spin interactions) allowing
to analyse their influence on the ground state behaviour.  Therefore,
strong interest still prevails in the experimental community to find
possible ways to synthesize a system which keeps the in-chain geometry
of Sr$_{2}$Cu(PO$_{4}$)$_{2}$, but with a smaller inter-chain coupling
which is still frustrating, along with an improved sample quality.

Formaly, K$_{2}$CuP$_{2}$O$_{7}$ belongs to the family of
M$_{2}$CuP$_{2}$O$_{7}$ (M=Li, Na, K) compounds, though the
arrangement of the magnetic CuO$_4$ building blocks is modified within
the family. The magnetic properties of Na$_{2}$CuP$_{2}$O$_{7}$ and
Li$_{2}$CuP$_{2}$O$_{7}$\ were reported earlier.\cite{nath2006a, alexander2007}
Na$_{2}$CuP$_{2}$O$_{7}$ and Li$_{2}$CuP$_{2}$O$_{7}$ both have an
in-chain exchange coupling $J_{1}$ of about 28 K and they undergo
magnetic ordering at around 5 K. Unlike the other members of this
family, our experiments reveal that K$_{2}$CuP$_{2}$O$_{7}$ shows
a much stronger in-chain exchange interaction ($J_{1} \simeq 141$ K)
but no ordering down to 2 K. Due to such a wide $T_{N}$--$J_{1}$
temperature range it can be considered as a very good candidate on
which to test the theoretical predictions of both static and dynamic
properties of an 1D HAF model.

In this paper, we present susceptibility, specific heat, $^{31}$P NMR
results and first principles band structure calculations on
K$_{2}$CuP$_{2}$O$_{7}$ in order to shed light on the 1D character of
the system. The manuscript is organized as follows: In the next
two sections, we describe the structural aspects, measurement
procedures and the theoretical methodology.  In section IV we
present the experimental results followed by the electronic structure
analysis. In the discussion section, we address the issues concerning
the almost ideal 1D nature of this compound inferred from the $^{31}$P
NMR and band structure calculations. At the end we give a short
summary and conclusions.
\begin{figure}[t]
\begin{center}
\includegraphics[%
  clip,
  width=8cm,
  angle=-0]{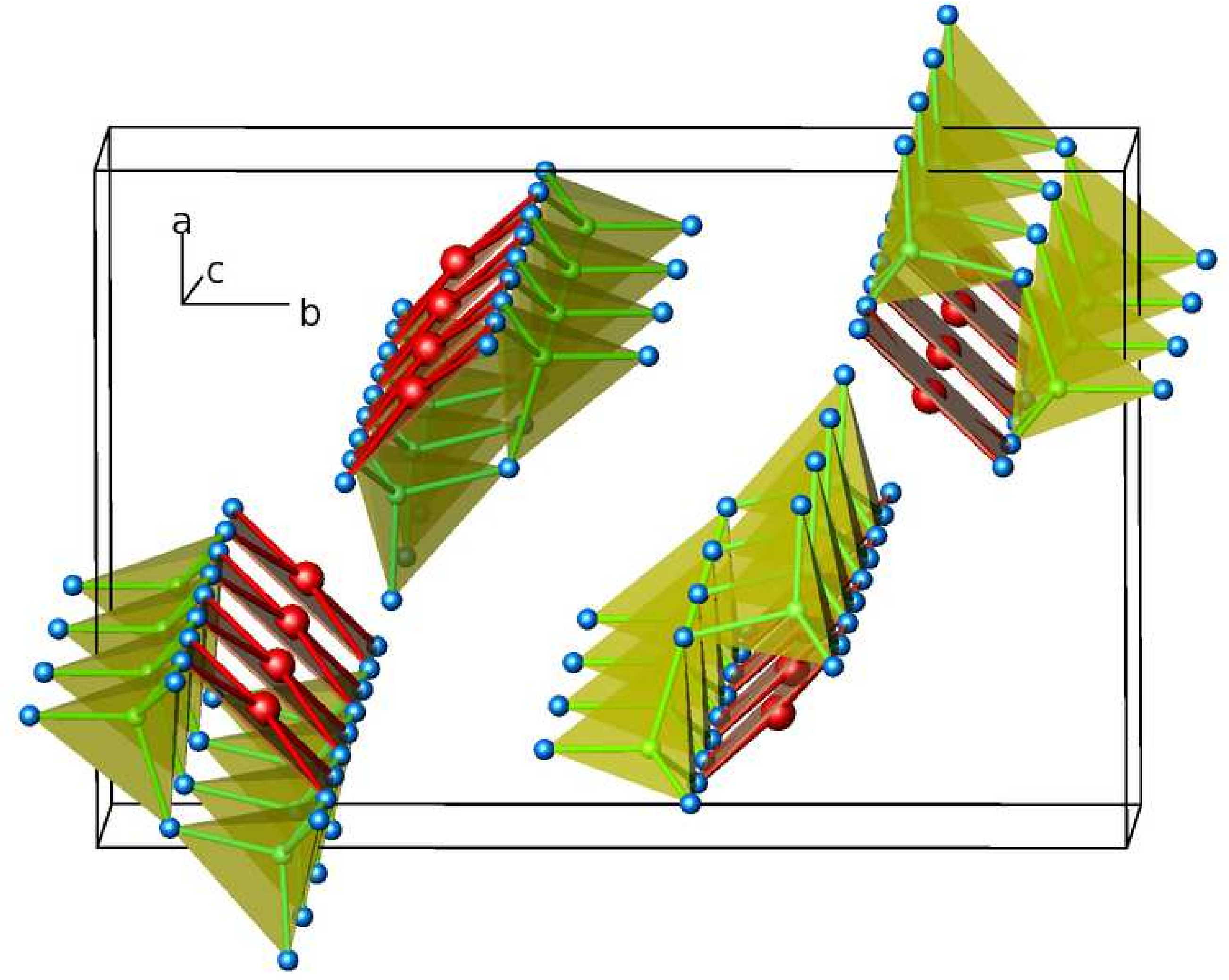}
\end{center}
\caption{\label{str}K$_{2}$CuP$_{2}$O$_{7}$ crystal structure.
The CuO$_{4}$ square planes (red) share their edges with the PO$_{4}$
tetrahedra (green) to form $\left[ \text{Cu(PO}_{\text{4}}\text{)}_{\text{2}}\right]
_{\infty }$ linear chains propagating along $c$-direction. The potassium
cations reside in between the chains (not shown here). }
\end{figure}

\section{Structure}

K$_{2}$CuP$_{2}$O$_{7}$ crystallizes in an orthorhombic unit cell with
space group $Pbnm$. The reported lattice constants are $a=9.509$ \AA ,
$b=14.389 $ \AA , and $c=5.276$ \AA .\cite{elmaadi1995} In the
crystal structure, isolated quasi planar CuO$_{4}$ units are linked by
PO$_{4}$ tetrahedra, forming [Cu(PO$_{4}$)$_{2} $]$_{\infty }$ chains
propagating along the crystallographic $c$-direction. A scheme of the
basic building blocks of the crystal is shown in Fig. \ref{str}.  The
super-exchange between Cu$^{2+}$ ions in K$_{2}$CuP$_{2}$O$_{7}$ will
be similar to that of edge sharing CuO$_{4}$ chains with every second
CuO$_{4}$ unit cut-off. In such a situation, $J_{2}$ (NNN) in the edge
shared system will become $J_{1}$ (NN) in our system (of the order of
$\approx$ 100 K), and $J_{2}$ in our system will be comparable to
$J_{4}$ (fourth neighbor interaction along the chain) in the edge
shared system, which is known to be quite negligible.  The chains are
well separated from each other since the potassium cations, K$^{1+}$
reside in between the chains.  The inter-chain interactions are
expected to be very weak, unlike Sr$_{2}$Cu(PO$_{4}$)$_{2}$ since the
chains do not lie in the same plane.  Magnetic properties of this
compound have not been reported yet.

\section{\textbf{Methods}}

\subsection{\textbf{Experimental}}

Polycrystalline K$_{2}$CuP$_{2}$O$_{7}$ was prepared by solid state
reaction techniques using K$_{2}$CO$_{3}$ ($99.9$\% pure), CuO
($99.99$\% pure) and NH$_{4}$H$_{2}$PO$_{4}$ ($99.9$\% pure) as
starting materials. The stoichiometric mixtures were fired at
$640$$^{\circ }$C for $60$ hours in air, with one intermediate
grinding. The samples were characterized using a STOE powder
diffractometer with a Cu target ($\lambda _{av}=1.54182$\AA ).  The
powder pattern evidenced single phase material. The lattice parameters
obtained using a least-square fit procedure are $a=9.541(3)$ \AA ,
$b=14.407(5)$ \AA , $c=5.253(1)$ \AA , close to the previously
reported values.\cite{elmaadi1995}

Magnetization ($M$) data were measured as a function of temperature $T$
between 2 K and 400 K in fields up to 5 T on powder samples in a
commercial (Quantum design) SQUID (superconducting quantum interference
device) magnetometer. Specific heat $C_{p}(T)$ measurements were performed
on a pressed pellet using relaxation method in a commercial PPMS
equipment(Quantum design). NMR measurements were carried out using pulsed
NMR techniques on $^{31}$P nuclei (nuclear spin $I=\frac{1}{2}$ and gyromagnetic
ratio $\gamma /2\pi $ = $17.237$ MHz/Tesla) at 70 MHz which corresponds to
an applied field of about 40.6 kOe. Spectra were obtained by Fourier
transform (FT) of the NMR echo signal using a $\pi /2$ pulse with width of
about $2$ $\mu s$. The NMR shift $K(T)=\left[ \nu \left( T\right) -\nu _{ref}
\right] /\nu _{ref}$ was determined by measuring the resonance frequency of
the sample ($\nu \left( T\right) $) with respect to a standard H$_{3}$PO$_{4}
$ solution (resonance frequency $\nu _{ref}$). The spin-lattice relaxation
rate $\left( 1/T_{1}\right) $ was measured by the saturation recovery
method. Our attempt to see the $^{63}$Cu signal was not successful due to the fast
relaxation at the magnetic site.

\subsection{\textbf{Theory}}

The bandstructure calculations presented here utilized version 5.00-18
of the full-potential local orbital band structure\cite{fplo1,fplo2}
(FPLO) method. The structure, lattice constants and atomic positions
have been taken from experiment \cite{elmaadi1995}.  The core states
have been treated fully-relativistically while the semi-core (K:
$3s3p$, Cu: $3s3p$, P: $2s2p$) and the valence states (K: $3d4s4p$,
Cu: $3d4s4p$, P: $3s3p3d$, O: $2s2p3d$) are treated
scalar-relativistically.  The extent of the valence basis functions
was optimized with respect to the total energy\cite{helmut}. The
Brillouin zone sampling was based on 216 $k$-points in the irreducible
wedge.  We have used the Perdew and Wang\cite{PW92} flavor of the
exchange and correlation potential when performing the calculations
within the local density approximation (LDA).

\section{\textbf{Results}}

\subsection{Susceptibility and specific heat}

\begin{figure}[b]
\begin{center}
\includegraphics[%
  clip,
  width=8cm,
  angle=-0]{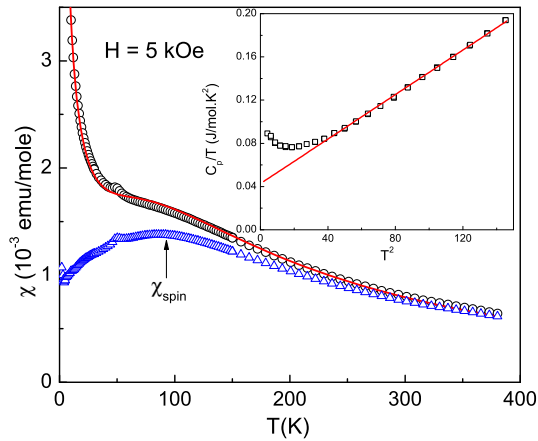}
\end{center}
\caption{\label{susceptibility}Magnetic susceptibility ($M/H
$) vs. temperature $T$ \ for K$_{2}$CuP$_{2}$O$_{7}$ (open circles)
in an applied field of
5 kOe. The solid line is best fit of the data to Eq. 1. Spin susceptibility
$\chi_{spin}$ is plotted (open triangles) after subtracting the
Curie contribution.  In the inset,
$C_{p}/T$ is plotted
as a function of $T^{2}$ for the low temperatures. Solid line represents a
linear fit.}
\end{figure}

Bulk magnetic susceptibility $\chi (T)$ (= $M/H$) was measured as a
function of temperature in an applied field of 5 kOe (Fig.
\ref{susceptibility}).  As shown in the figure, the sample exhibits
a shoulder at about 100 K, suggesting a maximum in this temperature
range, which is a hallmark of low-dimensional magnetic interactions.
With further decreasing temperature, $\chi (T)$ increases again in a
Curie-Weiss manner. Common sources for such an increase are the
presence of paramagnetic impurities or finite length chains due to
disorder. However, an intrinsic Curie like tail is expected in a
quasi 1D $S=\frac{1}{2}$ chain, if DM interaction and/or a staggered
$g$-factor anisotropy are present.\cite{oshikawa1997, affleck1999}
Among the 3d systems, Cu benzoate \cite{dender1996} and pyrimidine
Cu dinitrate\cite{feyerherm2000} are well known examples. No obvious
features associated with LRO were observed for $1.8$ K $\leq $ $T$
$\leq $ $400$ K.

In order to fit the bulk susceptibility data, we decomposed $\chi $ into

\textit{%
\begin{equation}
\chi =\chi _{0}+\frac{C_{imp}}{T+\theta_{imp} }+\chi _{spin}(T)  \label{Chi}
\end{equation}%
}where, the first term $\chi _{0}$ is temperature independent and accounts for the
 diamagnetism of the core electron shells  and Van-Vleck
paramagnetism  of the open shells of the Cu$^{2+}$ ions. The
second term $\frac{C_{imp}}{T+\theta_{imp} }$ is the low $T$ Curie-Weiss contribution due to
paramagnetic species in the sample. $\chi _{spin}(T)$ is the spin
susceptibility for a uniform $S=\frac{1}{2}$ 1D HAF system, which is known quite precisely
 over the whole measured temperature range. We took the
Johnston's expression\cite{johnston2000} valid for
$5\times 10^{-25} \leq \frac{T}{J_{1}}\leq 5$.

A fit of Eq. 1 to the experimental data leads to the parameters
$\chi _{0} \simeq $ 2.7$\times$10$^{-4}$ emu/mole, $C_{imp} \simeq $
0.026 emu.K/mole, $\theta_{imp} \simeq $ 0.7 K,  and
$J_{1}$ $\simeq$ 145 K.
Because of large Curie tail, such a fit is unstable at high temperatures.
Therefore we reduced the number of fitting
parameters by fixing $g$ = 2.2 (obtained from NMR
shift analysis)\cite{footnote1}.
%The slightly reduced value of $g$
%is attributed to the low-$T$ Curie contribution, while
The value of $\chi_0$ is comparable to that found in
Sr$_{2}$CuO$_{3}$\cite{motoyama1996}. The Curie contribution present
in the sample would correspond to a defect spin concentration of
$6.9$ \% assuming defect spin $S=\frac{1}{2}$. Alternatively, it
would correspond to the contribution expected for finite chains with
a length of $\approx$ 15 spins according to the calculation of
Schmidt {\it et al.,}\cite{schmidt2003}.  We add in Fig. 2 a plot of
$\chi_{spin}$ as a function of temperature, after subtracting the
Curie contribution. Now a pronounced broad maximum around 100 K is
well resolved. The small peak at 50 K corresponds to a parasitic
contribution of adsorbed oxygen . Unfortunately, we failed to
improve the quality of the sample, since further annealing (in
flowing Ar or vacuum) did not lead to any noticeable change in the
XRD pattern.

The information about the lattice dimensionality can be obtained from
the low temperature specific heat $C_{p}(T)$.  For a $S=\frac{1}{2}$
chain, one expects a linear term, whereas for a square lattice, one
expects a leading quadratic term in the absence of excitation gap.  In
the $C_{p}/T$ vs. $T^{2}$ plot (shown in the inset of
Fig. \ref{susceptibility}) in the temperature range 5 K $\leq T\leq$
12 K, one can see clearly that the plot follows a straight line over a
considerable temperature range. This indicates that $C_{p}(T)$ is a
sum of a linear and a cubic contribution. Since the cubic term
corresponds to the expected contribution of the phonons, this
demonstrate that the leading term of the magnetic contribution is
linear in $T$. For an $S=\frac{1}{2}$ HAF chain, theoretical
calculations\cite{klumper1998, johnston2000} predict for low
temperatures ($T<0.2J_{1}$);
$\frac{C_{p}}{T}=\frac{2R}{3J_{1}}=\gamma_{theo}$.  With $J_{1}\simeq
141$ K obtained from the NMR shift $K(T)$ analysis (presented later),
this corresponds to a value $\gamma _{theo}\simeq 0.04$ J/K$^{2}$
mole. Fitting the measured data in the range 5 K $\leq T\leq $ 12 K,
we obtained $\gamma _{\exp }\simeq 0.042$ J/K$^{2}$mole. The value of
$\gamma _{\exp }$ is close to the value predicted theoretically. This
strongly supports the quasi 1D nature of the spin system in
K$_{2}$CuP$_{2}$O$_{7}$. No other anomaly was observed down to $2$ K
in $C_{p}(T)$, suggesting the absence of magnetic order. Below $5$ K,
our experimental data deviates upward from the straight line fit
(inset Fig.~\ref{susceptibility}).  This is likely related to the
contributions of paramagnetic impurities and of chain ends, which
should follow a $C$ $\approx$ $A/T^{2}$ behavior.

\subsection{$^{31}$P NMR}
\begin{figure}[t]
\begin{center}
\includegraphics[%
  clip,
  width=8cm,
  angle=-0]{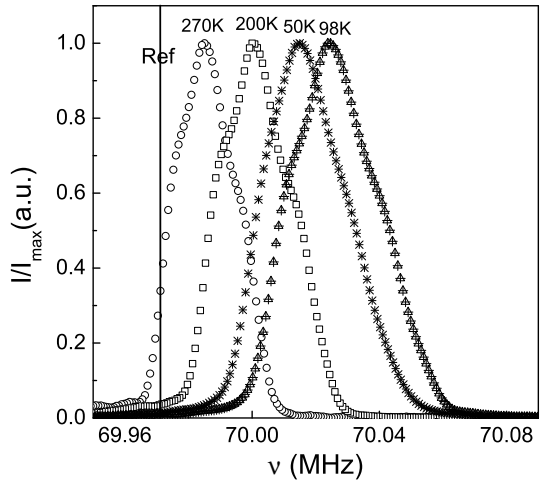}
\end{center}
\caption{\label{nmr} $^{31}$P NMR spectra at
different temperatures $T$. Solid line represents the nonmagnetic $^{31}$P
reference.}
\end{figure}

Although our analysis of the susceptibility and specific heat suggests
the presence of a $S = \frac{1}{2}$ 1D HAF system, the evidence is
only of preliminary nature because of the large Curie tail in $\chi
(T)$ at low $T$.  In order to gain a more reliable insight into this
system, we turned our attention towards NMR results.  NMR has the
advantage to be much less sensitive to contribution of defects or
impurities, because usually only the nucleus on an undistorted site
contributes to the narrow NMR line.  As shown in the crystal
structures (Fig. \ref{str}), K$_{2}$CuP$_{2}$O$_{7}$ has two
inequivalent $^{31}$P sites which are coupled inductively to Cu$^{2+}$
ions in the chain. Therefore $^{31}$P NMR can probe accurately the
low-lying excitations in the spin chain. Our $^{31}$P NMR spectra
consist of a single spectral line as is expected for $I=\frac{1}{2}$
nuclei (Fig. \ref{nmr}).  Although there are two inequivalent $^{31}$P
sites present the crystal structure, a single resonance line implies
that both the $^{31}$P sites in this compound are nearly
identical. With decreasing temperature, the NMR line shifts away from
the Larmor frequency, broadens paramagnetically but the overall line
shape remain same down to $2$ K. The asymmetric shape of the spectra
corresponds to a powder pattern due to an asymmetric hyperfine
coupling.
\begin{figure}[t]
\begin{center}
\includegraphics[%
  clip,
  width=8cm,
  angle=-0]{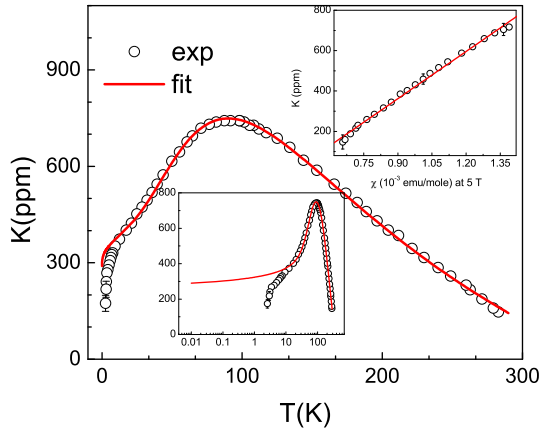}
\end{center}
\caption{\label{nmrshift} Temperature dependence of the $^{31}$P NMR
shift $K(T)$ of K$_{2}$CuP$_{2}$O$_{7}$. The solid line is the fit
with Eq. 2 in the temperature range, $9$ K $\leq T\leq 300$ K and
further extrapolated down to $0$ K. Lower inset shows $K$ vs. $T$ on a
logarithmic temperature scale for improved visualisation of the
low-$T$ data. In the upper inset, $K$ vs. $\chi$ is plotted with
temperature as an implicit parameter and the solid line is the linear
fit.}
\end{figure}

The temperature dependence of the NMR shift $K$
is shown in Fig. \ref{nmrshift}. With
decreasing temperature $K(T)$ increases paramagnetically, then passes
through a broad maximum at $110$ K, which is an indicative of short-range
correlations, and decreases again smoothly towards
low temperatures. Below $\frac{T}{J_{1}}\simeq 0.028$, $K(T)$ shows a
much steeper decrease towards zero. As mentioned before the NMR has
an advantage
over bulk susceptibility. One measures accurately the $\chi_{spin}$  by
NMR shift without suffering from the contribution from the free spins and
extrinsic foreign phases, which limits the accuracy of the bulk susceptibility
measurements. Therefore, it is more reliable to extract the magnetic
parameters from the temperature dependence of the NMR shift rather than from
the bulk susceptibility. The conventional scheme of correlating $K(T)$ and
$\chi (T)$ is to plot, $K$ vs. $\chi _{spin}$ with $T$ as
an implicit parameter. Then the slope
yields the average hyperfine coupling $A_{hf}$ between the $^{31}$P nucleus
and the two nearest-neighbor Cu$^{2+}$ ions.
In the case of K$_{2}$CuP$_{2}$O$_{7}$, because of the low $T$ Curie-tail
in $\chi (T)$ this $K(T)$ vs $\chi (T)$ plot shows a straight line only
for $T$ $>$ $110$ K (see upper inset of Fig.~\ref{nmrshift}.
Nevertheless, we can estimate
$A_{hf} \approx $ (4400$\pm$400) Oe/$\mu_{B}$, which is about two times
stronger than in other phosphate systems\cite{nath2005,nath2006}.
We then determine $J_{1}$ and $g$
by fitting the
temperature dependence of $K$ to the following equation,

\textit{%
\begin{equation}
K=K_{0}+\left( \frac{A_{hf}}{N\mu _{B}}\right) \chi _{spin}\left(
T,J_{1}\right)  \label{Shift}
\end{equation}%
}where $K_{0}$ is the temperature independent chemical shift. As
shown in Fig. \ref{nmrshift}, the $K(T)$ data fit rather well to Eq.
\ref{Shift} in the temperature range $9$ K $\leq $ $T$ $\leq $ $300$
K. Using $A_{hf}\simeq 4400$ Oe/$\mu _{B}$ (obtained from the $K$ vs
$\chi$ analysis) we obtained $K_{0}\simeq -890ppm$, $J_{1}\simeq
(141\pm 5)K $, and $g\simeq 2.2$. Below about $8$ K, $K(T)$ show a
significant deviation from the fit (see lower inset of
Fig.~\ref{nmrshift}) which shall be discussed later.
\begin{figure}[b]
\begin{center}
\includegraphics[%
  clip,
  width=8cm,
  angle=-0]{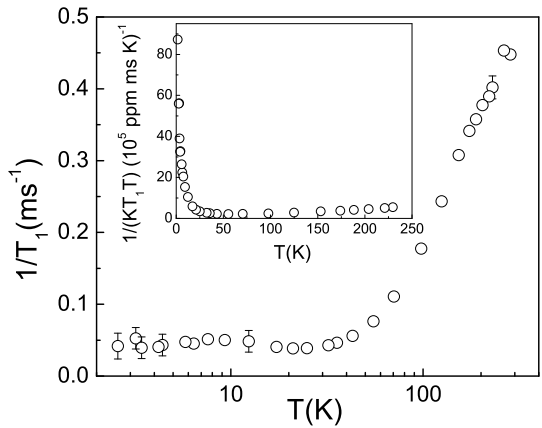}
\end{center}
\caption{\label{relax}
Spin-lattice
relaxation rate $1/T_{1}$ vs. temperature $T$ \ measured at 70 MHz. In the
inset, $1/(KT_{1}T) $ is shown all over the temperature range.}
\end{figure}

The temperature dependence of $^{31}$P $1/T_{1}$ is presented in
Fig. \ref{relax}. For a $I=\frac{1}{2}$ nucleus the recovery of the
longitudinal magnetization is expected to follow a single
exponential behavior. In the experiment, we indeed observed single
exponential behaviour down to $30$ K while below $30$ K it fitted
nicely to the stretch exponential with a reduced exponent. We didn't
observe any anomaly or divergence in $1/T_{1}(T)$ down to $2$ K
which indicates the absence of magnetic ordering.\ For $2$ K $ \leq
$ $T$ $\leq $ $30$ K, $1/T_{1}$ almost remains constant while for
$T$ $\geq 30$ K, it increases strongly with temperature. A slight
shoulder is visible in the plot $1/T_{1}$ versus $lnT$ around 180 K,
i.e just above $J_{1}$, suggesting a regime crossover in this
temperature range.
%linearly
%then shows a plateau at about 150K and again increases linearly with
%temperature but with a different slope.

\subsection{\textbf{First Principles and Tight Binding}}

\begin{figure}[t]
\begin{center}
\includegraphics[%
  clip,
  width=8.5cm,
  angle=-0]{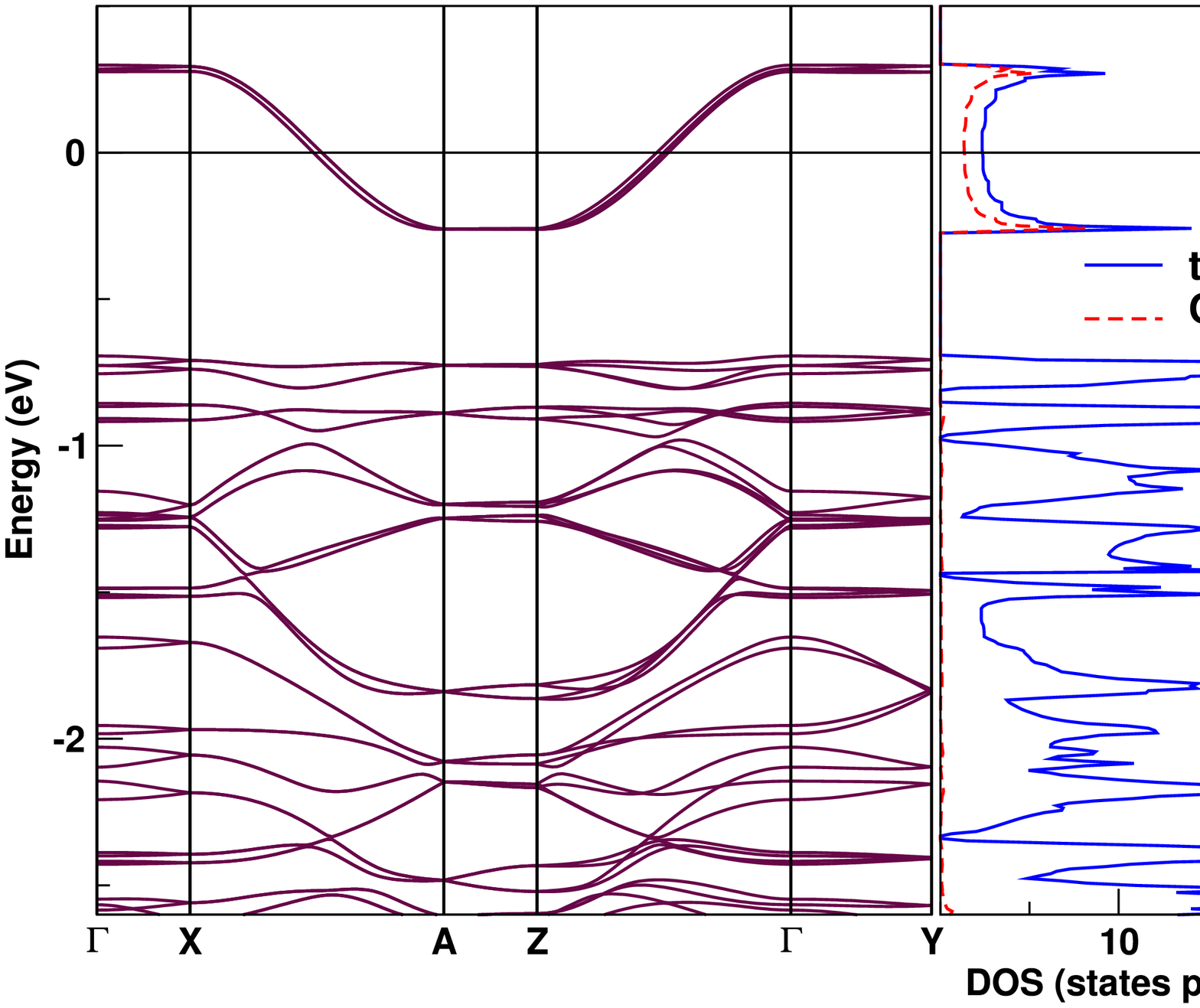}
\end{center}
\caption{\label{bands}(color online) {\bf Left panel:} The non-magnetic
band structure of
K$_{2}$CuP$_{2}$O$_{7}$ along the high-symmetry directions of a standard
orthorhombic unit cell. $X - A$ and $Z - \Gamma$ denote the directions along
the chain, ({\it i.e.} along crystallographic $c$ axis) wherein the
dispersion is the
largest while the perpendicular-to-chain directions $\Gamma - X$ and $\Gamma - Y$
are rather non-dispersive. There are four copper sites per unit cell leading
to four half-filled metallic bands at the Fermi level, well separated from
all other bands. {\bf Right panel:} The total DOS for
K$_{2}$CuP$_{2}$O$_{7}$ along with the $Cu - d_{x^{2}-y^{2}}$
% and
%$O - p_{\sigma}$
orbitally resolved contributions. Comparison with the
band structure clearly elucidates that the half-filled metallic band at the
Fermi level is primarily from  $Cu - d_{x^{2}-y^{2}}$ and
$O - p_{\sigma}$ molecular plaquette orbital.   }
\end{figure}

Collected in Fig. \ref{bands} are the non-magnetic band structure and
the density of states (DOS). The antibonding band is made from a
half-filled Cu - 3$d_{x^{2}-y^{2}}$ - O - 2$p_{\sigma}$ molecular
orbital belonging to the CuO$_{4}$ plaquettes. The band structure shows
strong dispersion of about 55 meV parallel to the chain direction
$\Gamma Z, XA$, but is nearly dispersion-less within the
crystallograhic $a-b$ plane ({\it i.e.} perpendicular to the direction
of the chains), indicative of the strong 1D character in this
system. Within LDA, we get a metallic behavior, though the system is
an insulator, suggested by its blue color.\cite{elmaadi1995} This is a
well known artifact of LDA wherein the effect of strong correlations
is under estimated. Including the strong correlations for the Cu 3$d$
states will open up the insulating gap. This can be achieved via two
possible ways: (a) performing an LDA+$U$ calculation
self-consistently, (b) mapping the results from LDA first to a
tight-binding model (TBM). The TBM is then mapped on to a Hubbard
model, and subsequently to a Heisenberg model because the system
belongs to the strong correlation limit $U$ $\gg$ t (t is the leading
transfer integral) at half filling.

\begin{figure}[H]
\begin{center}
\includegraphics[%
  clip,
  width=8cm,
  angle=-0]{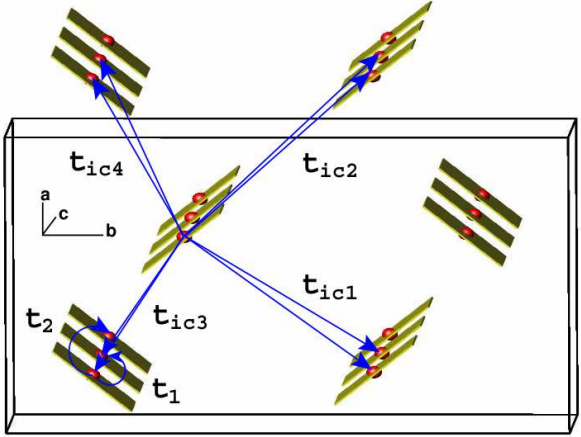}
\end{center}
\caption{\label{paths}(color online) The various hopping paths considered
in our tight-binding model to reproduce the half-filled metallic LDA band are
shown here. We have considered two hoppings along the chain ($t_{1},t_{2}$)
 and four
inter-chain hoppings ($t_{ic1},t_{ic2},t_{ic3},t_{ic4}$) in all. Starting
from one chain, Cu in the next neighbor chain are shifted by
half of the in-chain Cu-Cu distance, leading to two identical
interactions to two Cu-spins in each adjacent chains. }
\end{figure}

\begin{figure}[H]
\begin{center}
\includegraphics[%
  clip,
  width=8.5cm,
  angle=-0]{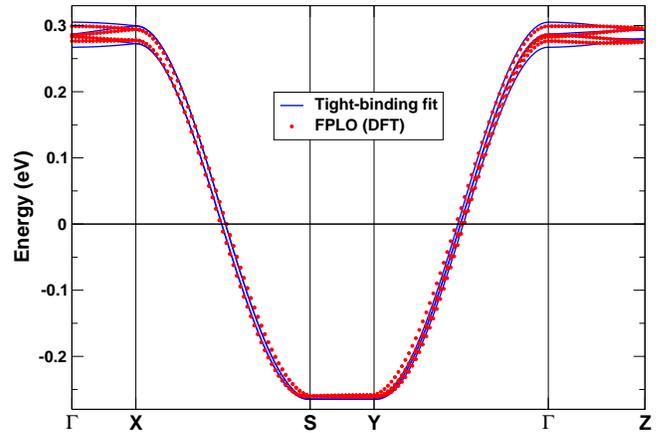}
\end{center}
\caption{\label{fit}(color online) The superposition of the total
band structure from the FPLO density functional theory (DFT) calculations
along with
the calculated eigenvalues from the four-site one-band TBM.
The TBM fit is quite consistent with the DFT results.}
\end{figure}

In order to better understand the microscopic magnetic interactions in
this system, we have followed the second option mentioned above.
Firstly we have considered the TBM,\\
\begin{equation}
H =  \sum_{\langle i,j \rangle,\sigma}
t_{ij}C^{\dagger}_{i,\sigma}C_{j,\sigma} + \sum_{i}\epsilon_{i}\hat{n_{i}}
\end{equation}
where $t_{ij}$ are the hopping integrals, $C^{\dagger}_{i,\sigma}C_{j,\sigma}$
are annihilation and creation operators. From the LDA band structure,
we have extracted only the 4 antibonding Cu - 3$d_{x^{2}-y^{2}}$ bands and
performed a fit to the TBM.
All the hopping paths considered in
our model are shown in Fig. \ref{paths}. The hopping integrals were
calculated using the steepest descent method.
The resulting parameters which
provided the best fit to the LDA band structure (Fig. \ref{fit})
are collected in Table~\ref{hoppings}.
The strength of the NN
hopping t$_1$ along the chain is two orders of magnitude
larger than all the other
hoppings, attributing the strong one-dimensionality to the interaction
of the plaquettes along the chain.
The individual exchange constants are
calculated using $J_{ij}^{AFM}$ = $4t^{2}_{ij}$/$U_{eff}$.  The value of $U_{eff}$
for K$_{2}$CuP$_{2}$O$_{7}$ is set to 4.5 eV, same as the choice in
related 1D compounds (Sr,Ba)$_{2}$Cu(PO$_{4}$)$_{2}$ \cite{johannes2006}.
The total exchange constant is given by $J^{total}$ = $J^{AFM}$ + $J^{FM}$.
The $J$ values collected in Table~\ref{hoppings} are indicative of the
$J^{AFM}$ only.
In K$_{2}$CuP$_{2}$O$_{7}$ the
CuO$_{4}$ plaquettes are separated from each other, with no corner-
or edge-sharing oxygens. The absence of direct connections between the
plaquettes largely suppresses ferromagnetic interactions between the
copper sites. The strength of the FM interactions was shown to be
very small in the related systems
(Sr,Ba)$_{2}$Cu(PO$_{4}$)$_{2}$ \cite{johannes2006}, and this result should
equally hold for our system.

\begin{center}
\begin{table}[H]
\begin{tabular*}{0.47\textwidth}%
     {@{\extracolsep{\fill}}|c|c|c|c|c|c|c|}
\hline
&  \textbf{$t_1$}   &  \textbf{$t_2$}   &   \textbf{$t_{ic1}$}   &   \textbf{$t_{ic2}$}  &   \textbf{$t_{ic3}$}   &   \textbf{$t_{ic4}$}    \tabularnewline
\hline
& & & & & & \tabularnewline
$$ (meV)  $$  & $$ 138 $$ & $$ 2 $$  & $$ 2 $$  & $$ 2 $$ & $$ 0.8 $$ & $$ 4 $$  \tabularnewline
& & & & & & \tabularnewline
\hline
\end{tabular*}

\vspace{0.5cm}

\begin{tabular*}{0.47\textwidth}%
     {@{\extracolsep{\fill}}|c|c|c|c|c|c|c|}
\hline
& & & & & & \tabularnewline
  &  \textbf{$J_{1}^{LDA}$}  &  \textbf{$J_{2}^{LDA}$}  &  \textbf{$J_{ic1}^{LDA}$}  &  \textbf{$J_{ic2}^{LDA}$}  &  \textbf{$J_{ic3}^{LDA}$}  &  \textbf{$J_{ic4}^{LDA}$}   \tabularnewline
& & & & & & \tabularnewline
\hline
& & & & & & \tabularnewline
$$ (K) $$ & $$ 196 $$ & $$ 0.04 $$ &  $$ 0.04 $$ & $$ 0.04 $$ & $$ 0.007 $$ & $$ 0.16 $$ \tabularnewline
& & & & & & \tabularnewline
\hline
\end{tabular*}
\caption{\label{hoppings}Hopping parameters (in meV) and the corresponding
exchange constants $J$ ( in K) from an effective one-band tight-binding model.
The hopping paths used are indicated in Fig. \ref{paths}.  }
\end{table}
\end{center}

\section{\textbf{Discussion}}

The quality of the fit for the NMR shift supports the presence of an $S=\frac{1}{2}$
HAF chain. The exchange coupling ($J_{1}$ $\simeq$ 141 K) is
comparable to that found in other
phosphates\cite{nath2005} and in nice agreement with the effective NN
super-exchange constant, $J_{1}^{LDA}$ = 196 K obtained from the TBM.
The slight over-estimation may stem from the fact that $U_{eff}$ is not exactly
known, along with some FM contributions. Such a over-estimation of $J$
by LDA is well known and is also observed in (Sr,Ba)$_{2}$Cu(PO$_{4}$)$_{2}$ \cite{johannes2006}.
The next nearest neighbor (NNN) super-exchange constant J$_{2}^{LDA}$ = 0.04 K,
is extremely small, so frustration coupling, if any should be negligible.
The
ratio of first and second neighbor in-chain coupling is
$J_{1}^{LDA}$/$J_{2}^{LDA}$
$\gtrsim$ 5000, the largest  found in $S$ = $\frac{1}{2}$ chain system.
The ratio of in-chain to the strongest (frustrating) inter-chain coupling is
$J_{1}^{LDA}$/$J_{ic4}^{LDA}$ $\gtrsim$ 1000, which is two orders of  magnitude larger
than in Sr$_{2}$Cu(PO$_{4}$)$_{2}$ ($J_{1}^{LDA}$/$J_{ic}^{LDA}$ $\sim$ 70) \cite{johannes2006} and one order
of magnitude larger than Sr$_{2}$CuO$_{3}$ ($J_{1}^{LDA}$/$J_{ic}^{LDA}$
$\sim$ 500) \cite{rosner1997,footnote} and  making K$_{2}$CuP$_{2}$O$_{7}$
an even better
realization of 1D HAF behavior than the Sr analogue.
We have estimated the N\'eel temperature  of
K$_{2}$CuP$_{2}$O$_{7}$ ($T_{N}^{K}$) by adapting a
simple mean field approximation\cite{rosner1997} and comparing it to
the Sr analogue ($T_{N}^{Sr}$). Assuming that the anisotropy are the same in
both compounds we can write according to Ref.~\onlinecite{rosner1997}
\begin{equation}
\frac{T_{N}^{K}}{T_{N}^{Sr}} \approx \frac{\sqrt{J_{1}^{K}J_{ic}^{K}}}{\sqrt{J_{1}^{Sr}J_{ic}^{Sr}}}
\end{equation}
where, $T_{N}^{Sr}$ $\approx$ 85 mK\cite{belik2005}.
% $J_{\parallel}$ is the in-chain NN
% exchange and $J_{\perp}$ is the interchain exchange.
Since the incompleteness in the mapping should be the same for both
the compounds, we can directly compare the values of $T_{N}$. We have
used $J_{1}^{Sr}$ = 187 K, $J_{ic}^{Sr}$ = 0.23 K (geometrical average
of the 2 inter-chain exchanges), $J_{1}^{K}$ = 196 K and $J_{ic}^{K}$
= $\sqrt[4]{J_{ic1}^{LDA}J_{ic2}^{LDA}J_{ic3}^{LDA}J_{ic4}^{LDA}}$ =
0.037K.  Plugging these values in the above equation we get
$T_{N}^{K}$ $\approx$ 37 mK, which is a factor of two smaller than the
Sr analogue. This value sets a upper bound, because the value of
$J_{ic}$ is calculated using a mean-field approximation.  The absence
of anomalies in $K(T)$, $1/T_{1}(T)$ and the invariant spectral shape
down to low temperatures rules out the possibility of LRO down to 2 K,
consistent with the theoretical prediction of an extremely low
$T_{N}$.  In K$_{2}$CuP$_{2}$O$_{7}$, each CuO$_{4}$ plaquettes in one
chain has two identical neighbors in each adjacent chains, with the
same exchange interactions. For a strong in-chain AFM exchange, this
leads to a complete frustration of the inter-chain interactions.
Quantum fluctuations are therefore enhanced here which suppresses
$T_{N}$ to lower values.

For a 1D $S=\frac{1}{2}$ HAF, theoretical calculations predict a weak
logarithmic decrease of $\chi (T)$ upon approaching $T$ = 0 K (see
Fig. \ref{nmrshift}).  The lower inset of Fig.~\ref{nmrshift} shows
clearly that the decrease we observe in $K(T)$ of
K$_{2}$CuP$_{2}$O$_{7}$ is much more pronounced .  The susceptibility
of a 1D $S = \frac{1}{2}$ HAF at $T$ = 0 is exactly
known\cite{johnston2000}, $\chi (T = 0) = \frac{g^{2}\mu
_{B}^{2}}{k_{B}(J_{1})\pi^{2}}$.  Then, $K(T)$ at zero temperature can
be written as $K_{theo}(T=0)=K_{0}+\left[ \frac{A_{hf}g^{2}\mu
_{B}}{k_{B}(J_{1})}\right] \times \frac{1}{\pi ^{2}}$.  Using the
parameters ($K_{0}$, $A_{hf}$, $g$, and $J_{1}$) determined from our
$K(T)$ analysis, $K_{theo}(T=0)$ was calculated to be $260$ ppm.
However, our experimental value at $2$ K is much lower, only $180$
ppm, and $K(T)$ is still decreasing steeply with $T$ at this
temperature.  Thus, at the quantitative level, the theoretically
predicted logarithmic term fails to describe our experimentaly
observed decrease. A similar feature has been found experimentally in
$^{17}$O NMR on Sr$_{2}$CuO$_{3}$ below $ T/J_{1}\simeq 0.015$
\cite{thurber2001} and $^{31}$P NMR on (Sr,Ba)$_{2}$Cu(PO$_{4}$)$_{2}$
below $T/J_{1}\simeq 0.003$.\cite{nath2005} In both the cases, it is
argued that the decrease is unrelated to the onset magnetic order or a
spin-Peierls transition. In K$_{2}$CuP$_{2}$O$_{7}$, we neither saw
any indication of ordering in $1/T_{1}(T)$ nor in $C_{p}(T)$.
Similarly there is no signature of exponential decrease (or singlet
ground state) observed in $1/T_{1}(T)$ as is expected for a
spin-Peierls transition.  One possibility to account for this drop in
$K(T)$ is the presence of DM interaction arising from the fact that
in K$_{2}$CuP$_{2}$O$_{7}$ there is no center of inversion
symmetry-relating two neighboring copper atoms along the chain.
 In the
presence of DM interaction, application of a magnetic field parallel
to the DM vector opens a gap $\Delta$ in the magnetic excitation
spectra. Since in a first approximation $\Delta$ increases with $B$ to
the power 2/3, it shall not be visible in the $B$ = 0 specific heat
data or the low field susceptibility shown in
Fig. \ref{susceptibility}. However, further experiments are needed to
confirm or discard this explanation.

In NMR valuable information on the dynamic of low-energetic spin
excitations can be gained from the analysis of the temperature
dependence of nuclear spin-lattice relaxation rates. Therefore, it
is essential to analyze $1/T_{1}(T)$ carefully which yields
information about the imaginary part of the dynamic susceptibility
$\chi \left( \mathbf{q,\omega }\right) $.  Thus $1/T_{1}$ should
include contributions from both the uniform $\left( q=0\right) $ and
staggered $\left( q=\pm \frac{\pi }{a}\right) $ spin fluctuations. A
theoretical analysis by Sachdev\cite{sachdev1994} shows that the
staggered component is dominant at low temperatures ($T\ll J_{1}$).
Indeed, the uniform component leads to $1/T_{1}\propto T$, while the
staggered component gives $1/T_{1}$ = constant. Monte Carlo
calculations by Sandvik supported the validity of these results over
an appropriate temperature range.\cite{sandvik1995} As shown in Fig.
\ref{relax},
%our results are in nice agreement with these
%theoretical predictions. Thus
our experimentally observed constant
$1/T_{1}$ at low temperatures ($2K\leq T\leq 30K$) suggests the
dominance of staggered fluctuations at low temperatures. The $^{31}$P
form factor for such systems is defined in Ref.~\onlinecite{nath2005}.
Since $^{31}$P is located symmetrically between the Cu ions, the
fluctuations are expected to be filtered out provided the hyperfine
couplings are equal. Moreover in this case we still have a
significant remnant contribution from $q=\pm \pi /a$ which plays a
dominant role at low temperatures. The possible origin of the
remnant staggered fluctuations could be the unequal hyperfine
couplings. In fact, such features have been previously observed in a
few other 1D $S=\frac{1}{2}$ HAF systems where $J_{1}\gg
T_{N}$.\cite {thurber2001,takigawa1996, takigawa1997, nath2005,
nath2006}
%At high temperature (30K-150K) the linear behaviour could
%be attributed to the uniform fluctuations. $1/T_{1}$ shows a
%deviation from the linearity at 150K which corresponds to the energy
%scale of the exchange coupling.
%Moreover this feature is not clear to us.
%Probably the measurements under different applied fields could help
%to resolve this issue if it is related to some other relaxation
%mechanisms such as,  spin
%diffusion\cite{kikuchi2001}.
The strong increase of $1/T_{1}$ with T above 30 K could be
attributed to the uniform fluctuations. When the dominant
contribution is from $q=0$, then one expects a constant
$1/(KT_{1}T)$. In K$_{2}$CuP$_{2}$O$_{7}$, for $T\geq 30$ K, we
rather observe a weak temperature dependency (inset of
Fig.~\ref{relax}) which might be due to some remanent contributions
from the staggered fluctuations and/or some additional relaxation
mechanisms. This weak temperature dependence as well as the
reduction of the slope in the plot $1/T_{1}$ versus $T$ above 180 K
might be related to spin diffusion as observed in
$\alpha$-VO(PO$_{3}$)$_{2}$ above $J_{1}$\cite{kikuchi2001}.

The relaxation rate due to staggered fluctuations can be calculated
following the prescription of Barzykin\cite{barzykin2001}. For the
purpose of comparison of theory with experiment he defined the
normalized dimensionless NMR spin-lattice relaxation rate at
low-temperature $\left( 1/T_{1}\right) _{norm}=\frac{ \hbar
J_{1}}{A_{th}^{2}T_{1}}\approx 0.3$, where $A_{th}$ is $A_{hf}\left(
2h\gamma /2\pi \right) $. Assuming the fluctuations to be correlated
because of the exchange $J_{1}$ along the chains, $1/T_{1}$ can be
written as $1/T_{1}=\frac{0.3}{\hbar J_{1}/A_{th}^{2}}$. Using this
expression, $\left( 1/T_{1}\right) $ at the $^{31}$P site was\
calculated to be about $129$ sec,$^{-1}$ whereas our experimental
value is 50 sec,$^{-1}$ in the $2$ K $\leq T$ $\leq $ $30$ K range.
The experimental value is about two times smaller than the
theoretical value, likely due to the effect of the geometrical form
factor. Further on a weak logarithmic increase in $1/T_{1}$ is
theoretically expected at low temperatures.\cite {barzykin2001}
 %In (Sr,Ba)2Cu(PO4)2 which are having almost the equal $J_{1}$ as
%K$_{2}$CuP$_{2}$O$_{7}$, such a logarithmic increase in $1/T_{1}$ has been
%observed only below 2K.
In the present case, our measurements were not done down to low
enough temperatures to rise this contribution above our experimental
error.

\section{\textbf{Conclusion}}

In conclusion, our experimental and theoretical studies of
K$_{2}$CuP$_{2}$O$_{7}$ demonstrate that this compound presents a very
uniform and strongly 1D $S=\frac{1}{2}$ HAF chain system.  Our NMR,
specific heat and susceptibility measurements show good agreement with
theoretical predictions for a 1D $S = \frac{1}{2}$ HAF chain.  Thus,
the temperature dependence of the NMR shift $K$ can be fitted in good
agreement with model calculations giving $J_{1} \simeq $ (141$\pm$5)
K. A calculation of the exchange interactions using a TBM fitted to
the results of {\it ab initio} LDA band structure calculations,
subsequently mapped onto a Heisenberg model, leads to a quite similar
value of $J_{1} \simeq $ 190 K and evidences extremely weak NNN
interactions as well as extremely weak and ``frustrated'' inter-chain
exchange of the order of $J^{ic} \sim$ 0.1 K, in
contrast to Sr$_{2}$Cu(PO$_{4}$)$_{2}$ with
$J^{ic} \sim$ 9 K\cite{johannes2006} and
to Sr$_{2}$CuO$_{3}$ ``non-frustrated''  $J^{ic} \sim$ 10 K\cite{rosner1997}.
The absence of any
evidence for magnetic order above 2 K in the experimental data
confirms the weakness of the inter-chain exchanges.
Using the TBM
results, a rough estimate of the Ne\'el temperature results in a value
of $T_{N} \sim$ 37 mK, only.
K$_{2}$CuP$_{2}$O$_{7}$ has the smallest in-chain as well as frustrating
smallest inter-chain exchanges which do not suppress quantum fluctuations,
thereby making this system  an even better example of a 1D $S = \frac{1}{2}$
HAF system than the compounds reported so far.

We further studied the magnetic fluctuations by analysing the
temperature dependence of $1/T_{1}$. At low temperatures $1/T_{1}$
remains constant,
% but increases linearly at high temperatures
in
reasonable agreement with Sachdev's predictions for 1D $S=\frac{1}{2}$
HAF system where relaxation is dominated by $q=\pm \pi /a$
% and $q=0$ components
at low $T$.
%and higher $T$, respectively.
Below 5 K, we
obtained a decrease of the NMR shift $K$ which is more pronounced than
that expected for a 1D $S = \frac{1}{2}$ HAF.  It's origin is not yet
clear, but might be due to DM interactions.

\begin{acknowledgments}
We thank to A. V. Mahajan, F. Haarmann for their critical suggestions
on NMR. We would like to acknowledge Alois Loidl for some preliminary
NMR measurements done in his group laboratory. D.K. and
H.R. acknowledge financial support from the `Emmy-Noether-program'
of the DFG.
\end{acknowledgments}


\begin{thebibliography}{99}
\bibitem{footnote1}Making $g$ as a free parameter gives a reduced
 value of about 1.8. Nevertheless in either case it does not lead to a
noticeable change in the value of $J_{1}$ ($\sim$ 130 K).
\end{thebibliography}
\end{document}